\documentclass[
               showpacs,preprintnumbers,superscriptaddress,amsmath,amssymb]{revtex4}
\usepackage{graphicx,amssymb,bm}
\usepackage{epsfig}
\usepackage{color}
\usepackage{bbm,verbatim}
\usepackage{subfigure}

\begin{document}

\newcommand{\bQe}{{\stackrel{\text{\tiny$\boldsymbol{\leftrightarrow}$}}
                 {\boldsymbol{Q}}^{\text{\raisebox{-1.8ex}{$\boldsymbol{(e)}$}}}}\mspace{-3mu}}
\newcommand{\bQec}{{\stackrel{\text{\tiny$\boldsymbol{\leftrightarrow}$}}
                 {\boldsymbol{Q}}^{\text{\raisebox{-1.8ex}{$\boldsymbol{(e)}*$}}}}\mspace{-1mu}}
\newcommand{\bQm}{{\stackrel{\text{\tiny$\boldsymbol{\leftrightarrow}$}}
                 {\boldsymbol{Q}}^{\text{\raisebox{-1.8ex}{$\boldsymbol{(m)}$}}}}\mspace{-3mu}}
\newcommand{\bQmc}{{\stackrel{\text{\tiny$\boldsymbol{\leftrightarrow}$}}
                 {\boldsymbol{Q}}^{\text{\raisebox{-1.8ex}{$\boldsymbol{(m)}*$}}}}\mspace{-1mu}}
\newcommand{\bQex}{\vec{\boldsymbol{Q}}^{\boldsymbol{(e)}}_x\mspace{-3mu}}
\newcommand{\bQey}{\vec{\boldsymbol{Q}}^{\boldsymbol{(e)}}_y\mspace{-3mu}}
\newcommand{\bQez}{\vec{\boldsymbol{Q}}^{\boldsymbol{(e)}}_z\mspace{-3mu}}
\newcommand{\bQmx}{\vec{\boldsymbol{Q}}^{\boldsymbol{(m)}*}_x\mspace{-1mu}}
\newcommand{\bQmy}{\vec{\boldsymbol{Q}}^{\boldsymbol{(m)}*}_y\mspace{-1mu}}
\newcommand{\bQmz}{\vec{\boldsymbol{Q}}^{\boldsymbol{(m)}*}_z\mspace{-1mu}}

\newcommand{\Oe}{\widehat{\boldsymbol{\Omega}}^{\boldsymbol{(e)}}}
\newcommand{\tq}{\widehat{\boldsymbol{\mathbbm{q}}}}

\newcommand{\av}[1]{\langle\mspace{1mu}{#1}\mspace{1mu}\rangle}
\newcommand{\bs}[1]{\vec{\boldsymbol{#1}}}

\newcommand{\bT}{\overset{\text{\tiny$\boldsymbol{\leftrightarrow}$}}{\boldsymbol{T}}}
\newcommand{\bI}{\overset{\text{\tiny$\,\boldsymbol{\leftrightarrow}$}}{\boldsymbol{I}}}

\newcommand{\bII}{{\stackrel{\text{\scriptsize$\boldsymbol{\leftrightarrow}$}}
                 {\boldsymbol{\mathbb{I}}}}}
\newcommand{\bTT}{{\stackrel{\text{\scriptsize$\boldsymbol{\leftrightarrow}$}}
                 {\boldsymbol{\mathbb{T}}}}}
\newcommand{\beps}{{\stackrel{\text{\tiny$\boldsymbol{\leftrightarrow}$}}
                 {\boldsymbol{\mathbb{\epsilon}}}}}
\newcommand{\doint}{\displaystyle\oint}

\newcommand{\bMM}[2]{{\stackrel{\text{\scriptsize$\boldsymbol{\leftrightarrow}$}}
                 {\boldsymbol{\mathbb{M}}}_{#1}^{\text{\raisebox{-1.8ex}{$#2$}}}}}
\newcommand{\bNN}[2]{{\stackrel{\text{\scriptsize$\boldsymbol{\leftrightarrow}$}}
                 {\boldsymbol{\mathbb{N}}}_{#1}^{\text{\raisebox{-1.8ex}{$#2$}}}}}
\newcommand{\bOO}[2]{{\stackrel{\text{\scriptsize$\boldsymbol{\leftrightarrow}$}}
                 {\boldsymbol{\mathbb{O}}}_{#1}^{\text{\raisebox{-1.8ex}{$#2$}}}}}
\newcommand{\bU}[1]{{\stackrel{\text{\scriptsize$\boldsymbol{\leftrightarrow}$}}
                 {\boldsymbol{\mathbb{U}}}^{\text{\raisebox{-1.8ex}{$#1$}}}}}

\newcommand{\be}{\,\vec{\boldsymbol{e}}}
\newcommand{\brp}{\vec{\boldsymbol{r}}^{\,\prime}}
\newcommand{\bh}{\,\vec{\boldsymbol{h}}}
\newcommand{\ft}[1]{\bs{\mathcal{#1}}}
\newcommand{\td}{\,\mbox{\rm d}}
\newcommand{\eg}{\,\mbox{\it e.g.,\;}}
\newcommand{\dint}{\displaystyle\int}
\newcommand{\dsum}{\displaystyle\sum}
\newcommand{\ddint}{\overset{\mspace{15mu}\infty}{\underset{\mspace{-20mu}-\infty}{\dint\!\!\!\dint}}}
\newcommand{\ddintl}[1]{\underset{\mspace{0mu}{#1}}{\dint\!\!\!\dint}}
\newcommand{\pd}[2]{\frac{\partial{#1}}{\partial{#2}}}
\newcommand{\pdi}[1]{\partial_{#1}}
\newcommand{\then}{\;\;\;\Longrightarrow\;\;\;}
\renewcommand{\-}{\text{\,\rule[0.55ex]{5mm}{.5pt}}\,}
\renewcommand{\.}{\boldsymbol{\cdot}}
\newcommand{\x}{\boldsymbol{\times}}
\newcommand{\wz}{\widetilde{z}}
\newcommand{\tc}[1]{{\;\stackrel{(#1)}{\text{\raisebox{-0.5ex}{$\boldsymbol{\cdot\cdot}$}}}\;}}
\newcommand{\tcj}[1]{{\,\stackrel{(#1)}{\text{\raisebox{-0.5ex}{$\boldsymbol{:}$}}}\,}}
\newcommand{\tcc}[1]{{\,%
                     \text{\raisebox{-1.0ex}{$\stackrel{(#1)}{\stackrel{\text{\small\raisebox{-0.3ex}{$\boldsymbol{:}$}}}{\times}}$}}%
                     \;}}

\title{Decomposition of optical force 
into conservative and nonconservative components}

\author{Yikun Jiang}
\affiliation{State Key Laboratory of Surface Physics and Department of Physics, Fudan University, Shanghai 200433, China}
\affiliation{Key Laboratory of Micro and Nano Photonic Structures (Ministry of Education), Fudan University, Shanghai 200433, China}
\author{Jun Chen}
\affiliation{Institute of Theoretical Physics and Collaborative Innovation Center of Extreme Optics, Shanxi University, Shanxi, China }
\author{Jack Ng}
\email[]{jacktfng@hkbu.edu.hk}
\affiliation{Department of Physics and Institute of Computational and Theoretical Studies, Hong Kong Baptist University, Hong Kong}
\author{Zhifang Lin}
\email[]{phlin@fudan.edu.cn}
\affiliation{State Key Laboratory of Surface Physics and Department of Physics, Fudan University, Shanghai 200433, China}
\affiliation{Key Laboratory of Micro and Nano Photonic Structures (Ministry of Education), Fudan University, Shanghai 200433, China}
\affiliation{Collaborative Innovation Center of Advanced Microstructures, Nanjing University, Nanjing 210093, China}

\begin{abstract}
We present 
a multipole expansion theory for optical force exerting on a particle 
immersed in generic monochromatic free-space optical field.
Based on the theory, we have, for the first time,
successfully decomposed the optical force on a spherical particle of arbitrary size
into a conservative and a nonconservative parts, which are, respectively, written as a gradient of
a scalar function and curl of a vector function in an explicit and analytical form.
As a result, a scalar potential and a vector potential can be defined, up to gauge freedoms,
for the optical force.
The decomposition shed light on the understanding of the optical force and pave
a new way to engineer optical force for various purposes such as equilibrium
statistical mechanics as well as optical micromanipulation.
\end{abstract}


\maketitle

\section{Multipole expansion of optical force}

Physically, the period-averaged optical force acting on a particle in any monochromatic optical field
is written in two terms, the interception (extinction) force $\av{\bs F_{\text{int}}}$ and recoil force $\av{\bs F_{\text{rec}}}$.
Based on the T-matrix method \cite{bohren,tmatrix}, the multipole field theory \cite{rose,stratton}, and the irreducible tensor theory \cite{silver}, the optical force can be
written in terms of the electric and magnetic multipoles of various orders induced on the particle. To be specific,
the interception force $\av{\bs F_{\text{int}}}$ is given by
\begin{subequations}\label{Fmix}
\begin{eqnarray}
\av{\bs F_{\text{int}}}&=&\displaystyle\sum_{l=1}^{\infty} \av{\bs F^{(l)}_{\text{int}}}, \label{Fmixa} \\[1mm]
\av{\bs F^{(l)}_{\text{int}}}
&=&\av{\bs F^{\text{\,e}(l)}_{\text{int}}}+\av{\bs F^{\text{\,m}(l)}_{\text{int}}} \label{Fmixb} \\[3mm]
\av{\bs F^{\text{\,e}(l)}_{\text{int}}}&=&\dfrac1{2\,l!}\;\text{Re}\Bigl[(\nabla^{(l)} \bs{E}^*)
                          \tc{l} \bOO{\text{elec}}{(l)}\Bigr]  \label{Fmixe} \\[1mm]
\av{\bs F^{\text{\,m}(l)}_{\text{int}}}&=&\dfrac1{2\,l!}\;\text{Re}\Bigl[(\nabla^{(l)} \bs{B}^*)
                          \tc{l} \bOO{\text{mag}}{(l)}\Bigr]  \label{Fmixm}
\end{eqnarray}
\end{subequations}
and the recoil force $\av{\bs F_{\text{rec}}}$ reads
\begin{subequations} \label{Fsca}
\begin{eqnarray}
\av{\bs F_{\text{rec}}}&=&\displaystyle\sum_{l=1}^{\infty} \av{\bs F^{(l)}_{\text{rec}}},  \label{Fscaa} \\[1mm]
\av{\bs F^{(l)}_{\text{rec}}}
&=&\av{\bs F^{\text{\,e}(l)}_{\text{rec}}}+\av{\bs F^{\text{\,m}(l)}_{\text{rec}}} + \av{\bs F^{\text{\,x}(l)}_{\text{rec}}}  \label{Fscab} 
\end{eqnarray}
\begin{eqnarray}
\av{\bs F^{\text{\,e}(l)}_{\text{rec}}}
&=&-\dfrac{1}{4\pi\varepsilon_0}\,\dfrac{(l+2)\,k^{2l+3}}{(l+1)!\,(2l+3)!!}\;\text{Im}
\Bigl[\,\bOO{\text{elec}}{(l)\,\text{\raisebox{0.5ex}{$*$}}}\tc{l}  \bOO{\text{elec}}{(l+1)}\Bigr]  \label{Fscae} \\[1mm]
\av{\bs F^{\text{\,m}(l)}_{\text{rec}}}
&=&-\dfrac{\mu_0}{4\pi}\,\dfrac{(l+2)\,k^{2l+3}}{(l+1)!\,(2l+3)!!}\;\text{Im}
\Bigl[\, \bOO{\text{mag}}{(l)\,\text{\raisebox{0.5ex}{$*$}}}\tc{l}  \bOO{\text{mag}}{(l+1)} \,\Bigr] \label{Fscam} \\[1mm]
\av{\bs F^{\text{\,x}(l)}_{\text{rec}}}
&=&\dfrac{Z_0}{4\pi}\,\dfrac{k^{2l+2}}{l\;l!\,(2l+1)!!}\;\,\text{Re}\bigl[\,\bOO{\text{elec}}{(l)}\tc{l-1} \bOO{\text{mag}}{(l)\,\text{\raisebox{0.5ex}{$*$}}}\,\bigr]\tc{2} \beps,
 \label{Fscax}
\end{eqnarray}
\end{subequations}
where $k$ is the wave number in the transparent (lossless) background medium where the particle is located,
the quantities $\bs E$ and $\bs B$ denote, respectively, the incident electric and magnetic fields where the particle is immersed,
$\beps$ is the Levi-Civita tensor, whose components $\epsilon_{ijk}$ are antisymmetric with respect to the permutation of any pair of indices, and
the superscripts ``e'', ``m'', and ``x'' denote, respectively, the contribution due to
the electric multipoles, magnetic multipoles and hybrid term.
The electric and magnetic multipoles of order $l$ are described by
totally symmetric and traceless \cite{apple} rank-$l$ tensors $\bOO{\text{elec}}{(l)}$ and $\bOO{\text{mag}}{(l)}$, respectively.
The lower order cases with $l = 1$, 2, 3, 4, and 5 correspond
to the dipole, quadrupole, octupole, hexadecapole, and
dotriacontapole, respectively.
The multiple contraction between two tensors of ranks $l$ and $l'$, denoted by \  $\!\tc{m}\!$,  \
yields a tensor of rank $l+l'-2m$ given by
\begin{equation}\label{AtcB}
\stackrel{\text{\scriptsize$\boldsymbol{\leftrightarrow}$}}
         {\boldsymbol{\mathbb{A}}}^{\text{\raisebox{-1.8ex}{$(l)$}}}
         \tc{m}
\stackrel{\text{\scriptsize$\boldsymbol{\leftrightarrow}$}}
         {\boldsymbol{\mathbb{B}}}^{\text{\raisebox{-1.8ex}{$(l')$}}}=
\mathbb{A}^{(l)}_{i_1\,i_2\,\cdots\,i_{l-m}\,\textcolor{blue}{{k_1\,k_2\,\cdots\,k_{m-1}\,k_m}}}\,
\mathbb{B}^{(l')}_{\textcolor{blue}{{k_m\,k_{m-1}\,\cdots\,k_2\,k_1}}\,j_{m+1}\,\cdots\,j_{l'-1}\,j_{l'}}, \quad 0\le m\le\min\,[\,l,l'\,],
\end{equation}
with the summation over repeated indices assumed. Here the tensor contraction is made consecutively over two nearest indices in two index sequences.
Equations (\ref{Fmix}) and (\ref{Fsca}) imply that
the interception force is simply the coupling of the electric multipoles
with the external electric field,  governed by Eq.~(\ref{Fmixe}), plus the coupling between the
magnetic multipoles and the external magnetic field, determined by Eq.~(\ref{Fmixm}).
The recoil force originates from the coupling between the electric multipoles of adjacent orders, denoted by Eq.~(\ref{Fscae}),
the magnetic multipoles of adjacent orders, delineated by Eq.~(\ref{Fscam}),
and the electric and magnetic multipoles of the same order, depicted by Eq.~(\ref{Fscax}).
Equations (\ref{Fmix}) and (\ref{Fsca}) constitute 
the complete multipole expansion of optical force on any particle, up to arbitrary orders of multipoles.

In the following, we give, for a spherical particle, the electric (magnetic) $2^l$-pole moments, $\bOO{\text{elec}}{(l)}$ \,[$\bOO{\text{mag}}{(l)}$],
in terms of the Mie coefficients $a_l$ ($b_l$) \cite{bohren} and the multiple gradient of the incident electric (magnetic) field.
For a general (non-spherical) particle, the multipole moments are determined by its T-matrix \cite{tmatrix},
instead of the simple Mie coefficients.
The T-matrix for a general non-spherical scatterer can only be evaluated numerically.

Define two totally symmetric tensors of rank $n$ by
\begin{equation}\label{Omega}
\begin{array}{rlll}
\bMM{\text{elec}}{(n)}&=\mathcal{\hat S}\bigl[\,\nabla^{(n-1)} \bs{E}\,\bigr]
=\mathcal{\hat S}\bigl[\,\overbrace{\nabla\,\nabla\cdots\nabla}^{n-1} \bs{E}\,\bigr] 
=\dfrac1n\,\dsum_{j=1}^{n}\partial_{i_1}\cdots\partial_{i_{j-1}}\partial_{i_n}\partial_{i_{j+1}}\cdots\partial_{i_{n-1}}\,E_{i_j} \\[4mm]
\bMM{\text{mag}}{(n)}&=\mathcal{\hat S}\bigl[\,\nabla^{(n-1)} \bs{B}\,\bigr]
=\mathcal{\hat S}\bigl[\,\overbrace{\nabla\,\nabla\cdots\nabla}^{n-1} \bs{B}\,\bigr] 
=\dfrac1n\,\dsum_{j=1}^{n}\partial_{i_1}\cdots\partial_{i_{j-1}}\partial_{i_n}\partial_{i_{j+1}}\cdots\partial_{i_{n-1}}\,B_{i_j}
\end{array}
\end{equation}
where $\nabla^{(n)}$ means taking $n$ consecutive gradients on a scalar or vector fields, whereas
$\mathcal{\hat S}$ denotes the symmetrizing operator,
$E_j$ ($B_j$) is the $j$-th Cartesian component of the electric (magnetic) field,
and $\pdi{i}$ represents the partial derivative $\displaystyle\pd{\ }{x_i}$ with respect to the $i$-th Cartesian coordinate.
The tensors $\bMM{\text{elec\,(mag)}}{(n)}$ so defined are obviously totally symmetric
(invariant under the permutation of any pair of indices), but they are not totally traceless
(vanishing under the contraction of any pair of indices) \cite{apple}.
To obtain totally traceless tensors for the electric and magnetic multipoles, we proceed to define another two (electric and magnetic)
sets of totally symmetric rank-$l$ tensors by
\begin{equation}
\bNN{\text{elec\,(mag)}}{(l,\,m)}=\mathcal{\hat S}\Bigl[\,
\overbrace{\bII\otimes\bII\otimes\cdots\bII\otimes}^{m}\; \bMM{\text{elec\,(mag)}}{(l-2m)}\,\Bigr] \quad\;\;
\text{with}\quad
\bNN{\text{elec\,(mag)}}{(l,\,0)}=\bMM{\text{elec\,(mag)}}{(l)},
\end{equation}
where $\bII$ denotes the unit dyad of dimension 3, and the symbol $\otimes$ represents the tensor product so that
the term in the square brackets means $m$ consecutive times of tensor product, resulting in
a rank-$l$ tensor independent of $m$. In component form, it reads
$$
\bNN{i_1i_2,\cdots\,i_l}{(l,\,m)}=\mathcal{\hat S}\bigl[\,\delta_{i_1i_2}\delta_{i_3i_4}\cdots\delta_{i_{l-2m-2}i_{l-2m-1}}
\bMM{i_{l-2m+1}i_{l-2m+2},\cdots\,i_l}{(l-2m)}\,\bigr].
$$
It can be proved that
\begin{equation}\label{nMnN}
\begin{array}{lll}
\bs n\x\Bigl[\,
\bs{n}^{(l-2m-1)}
\tc{l-2m-1}
\bMM{\text{elec\,(mag)}}{(l-2m)}\Bigr]
&=\dfrac{l}{l-2m}\,\bs n\x
\Bigl[\,
\bs{n}^{(l-1)} 
\tc{l-1}
\bNN{\text{elec\,(mag)}}{(l,\,m)} \Bigr],
\end{array}
\end{equation}
where $\bs n=\bs r/r$ is the unit vector in radial direction,
and $\bs n^{(m)}$ 
denotes a rank-$m$ tensor resulting from the tensor product of $m$ vectors $\bs n$.
For instance, $\bs n^{(0)}=1$, \ $\bs n^{(1)}=\bs n$, \ $\bs n^{(2)}=\bs n\otimes\bs n$, and $\bs n^{(3)}=\bs n\otimes\bs n\otimes\bs n$.
The totally symmetric and traceless multipole moments, $\bOO{\text{elec}}{(l)}$ and $\bOO{\text{mag}}{(l)}$,
which are therefore
usually referred to as $2^l$-pole moments, 
are derived based on theory of multipole fields \cite{rose}
\begin{equation}\label{Ol}
\begin{array}{ll}
\bOO{\text{elec\,(mag)}}{(l)}=\gamma^{(l)}_{\text{elec\,(mag)}}\,\dsum^{\lfloor\frac{l-1}2\rfloor}_{m=0}
d_{l,m}\,k^{2m}\,
\bNN{\text{elec\,(mag)}}{(l,\,m)}, \\[5mm]
d_{l,m}
=\dfrac{1}{4^m}\,\dfrac{l!}{m!}\,\dfrac{\Gamma(l-m+\frac12)}{\Gamma(l+\frac12)\,\Gamma(l-2m)}\,\dfrac{1}{l},
\mspace{50mu} \text{with} \quad d_{l,0}=1,
\end{array}
\end{equation}
where $\lfloor x\rfloor$ gives the greatest integer less than or equal to $x$ and
$\Gamma(x)$ denotes the Gamma function.
The lower order cases with $l = 1$, 2, 3, 4, and 5 correspond
to the dipole, quadrupole, octupole, hexadecapole, and
dotriacontapole moments, respectively.
For instance, $\bOO{\text{elec}}{(1)}$ reduces to the electric dipole moment $\bs p$,
$\bOO{\text{mag}}{(1)}$ is the magnetic dipole moment $\bs m$,
$\bOO{\text{elec}}{(2)}$ represents the electric quadrupole moment $\bQe$, and
$\bOO{\text{mag}}{(2)}$ delineates the magnetic quadrupole moment $\bQm$,
just to list a few lowest order cases.
The electric and magnetic polarizabilities, $\gamma_{\text{elec}}^{(l)}$ and $\gamma_{\text{mag}}^{(l)}$, 
depend on the Mie coefficients \cite{bohren} $a_l$ and $b_l$ of a spherical particle through
\begin{equation}
\gamma_{\text{elec}}^{(l)}=\dfrac{4l(2l+1)!!}{(l+1)}\;\dfrac{i\pi\varepsilon_0\,a_l}{k^{2l+1}},\qquad
\gamma_{\text{mag}}^{(l)}=\dfrac{4l(2l+1)!!}{(l+1)}\;\dfrac{i\pi\, b_l}{\mu_0k^{2l+1}}.
\end{equation}
In the Syst\`em International d'Unit\'es (SI), the dimensions of
$\gamma_{\text{mag}}^{(l)}$ and $\gamma_{\text{elec}}^{(l)}$ differ by a factor of $c^2$, viz,
$\gamma_{\text{mag}}^{(l)}/\gamma_{\text{elec}}^{(l)}=c^2\, (b_l/a_l)$.

\section{Decomposition into gradient and curl forces}

To decompose the optical force acting on a spherical particle
into a sum of an irrotational term of zero curl and a solenoidal (divergenceless) term of zero divergence,
or, equivalently, the conservative and non-conservative parts,
it is convenient to write the optical force in terms
of the multiple gradients $\bMM{\text{elec}}{(n)}$ and $\bMM{\text{mag}}{(n)}$ of the electric and magnetic fields defined in
(\ref{Omega}), 
instead of multipole moments $\bOO{\text{elec}}{(l)}$ and $\bOO{\text{mag}}{(l)}$. After lengthy algebra, the results read
\begin{subequations}\label{Fmixl}
\begin{eqnarray}
\av{\bs F^{\text{\,e}(l)}_{\text{int}}}
&\!=&\!\dfrac1{2\,l!}\; \text{\textcolor{red}{Re}}
 \dsum^{\lfloor\frac{l-1}2\rfloor}_{m=0}
 c_{l,m}k^{4m}\gamma^{(l)}_{\text{elec}}
     \bigl[\nabla^{(l-2m)}\bs{E}^*\,\bigr]
         \tc{l-2m}\bMM{\text{elec}}{(l-2m)},
         \label{Fmixle}\\
\av{\bs F^{\text{\,m}(l)}_{\text{int}}}
&\!=&\!\dfrac{1}{2\,l!}\;  \text{\textcolor{red}{Re}}
\dsum^{\lfloor\frac{l-1}2\rfloor}_{m=0}
 c_{l,m}k^{4m}\gamma^{(l)}_{\text{mag}}
    \bigl[\nabla^{(l-2m)}\bs{B}^*\,\bigr]
          \tc{l-2m}\bMM{\text{mag}}{(l-2m)},
          \label{Fmixlm}
\end{eqnarray}
\end{subequations}
\begin{subequations}\label{Fscal}
\begin{eqnarray}
\av{\bs F^{\text{\,e}(l)}_{\text{rec}}}
&\!=&\!
-\dfrac{1}{4\pi\varepsilon_0}\,\dfrac{(l+2)\,k^{2l+3}}{(l+1)!\,(2l+3)!!}
\; \text{\textcolor{blue}{Im}}
\dsum_{m=0}^{\lfloor\frac{l-2}2\rfloor}
g_{l,m}\, k^{4m+2}\,
\eta^{(l)}_{\text{elec}}
      \,\bMM{\text{elec}}{(l-2m)\,\text{\raisebox{0.5ex}{$*$}}}\tc{l-2m-1}   \bMM{\text{elec}}{(l-2m-1)}
\nonumber \\ &&\mspace{-7mu}
-\dfrac{1}{4\pi\varepsilon_0}\,\dfrac{(l+2)\,k^{2l+3}}{(l+1)!\,(2l+3)!!}
\; \text{\textcolor{blue}{Im}}
\dsum_{m=0}^{\lfloor\frac{l-1}2\rfloor}
f_{l,m}\, k^{4m}\,
\eta^{(l)}_{\text{elec}}
      \,\bMM{\text{elec}}{(l-2m)\,\text{\raisebox{0.5ex}{$*$}}}\tc{l-2m}   \bMM{\text{elec}}{(l-2m+1)}, \label{Fscale} \\
\av{\bs F^{\text{\,m}(l)}_{\text{rec}}}
&\!=&\!
-\dfrac{\mu_0}{4\pi}\,\dfrac{(l+2)\,k^{2l+3}}{(l+1)!\,(2l+3)!!}
\; \text{\textcolor{blue}{Im}}
\dsum_{m=0}^{\lfloor\frac{l-2}2\rfloor}g_{l,m}\, k^{4m+2}\,
   \eta^{(l)}_{\text{mag}}
      \,\bMM{\text{mag}}{(l-2m)\,\text{\raisebox{0.5ex}{$*$}}}\tc{l-2m-1}   \bMM{\text{mag}}{(l-2m-1)}
\nonumber \\ && \mspace{-7mu} {}
-\dfrac{\mu_0}{4\pi}\,\dfrac{(l+2)\,k^{2l+3}}{(l+1)!\,(2l+3)!!}
\; \text{\textcolor{blue}{Im}}
\dsum_{m=0}^{\lfloor\frac{l-1}2\rfloor}
f_{l,m}\, k^{4m}\,
\eta^{(l)}_{\text{mag}}
      \,\bMM{\text{mag}}{(l-2m)\,\text{\raisebox{0.5ex}{$*$}}}\tc{l-2m}\bMM{\text{mag}}{(l-2m+1)},
       \label{Fscalm}
\end{eqnarray}
\begin{eqnarray}
\av{\bs F^{\text{\,x}(l)}_{\text{rec}}}
&\!=&\!\dfrac{Z_0}{4\pi}\,\dfrac{k^{2l+2}}{l\;l!\,(2l+1)!!}
\;\text{\textcolor{red}{Re}}
\dsum_{m=0}^{\lfloor\frac{l-1}2\rfloor}
h_{l,m}\, k^{4m}\eta^{(l)}_{\text{hyb}}\,\bigl[\,\bMM{\text{elec}}{(l-2m)}\tc{l-2m-1}
 \bMM{\text{mag}}{(l-2m)\,\text{\raisebox{0.5ex}{$*$}}}\,\bigr]\! \tc{2} \beps,  \label{Fscalx}
\end{eqnarray}
\end{subequations}
where $\eta^{(l)}_{\text{elec}}=\gamma^{(l+1)}_{\text{elec}}\gamma^{(l)\,*}_{\text{elec}}\,$, \
$\eta^{(l)}_{\text{mag}}=\gamma^{(l+1)}_{\text{mag}}\gamma^{(l)\,*}_{\text{mag}}\,$, \ and
$\eta^{(l)}_{\text{hyb}}=\gamma^{(l)}_{\text{elec}}\gamma^{(l)\,*}_{\text{mag}}\,$, are products of polarizabilities.
In deriving Eqs.~(\ref{Fmixl}) and (\ref{Fscal}), we have used the following mathematical identities
\begin{subequations}
\begin{equation}
\begin{array}{lll}
\bOO{\text{elec\,(mag)}}{(l)\,\text{\raisebox{0.5ex}{$*$}}}\tc{l}  \bOO{\text{elec\,(mag)}}{(l+1)}
&=
\dsum_{m=0}^{\lfloor(l-1)/2\rfloor}
f_{l,m}\, k^{4m}\, \eta^{(l)}_{\text{elec\,(mag)}}\,
\bMM{\text{elec\,(mag)}}{(l-2m)\,\text{\raisebox{0.5ex}{$*$}}}\tc{l-2m}   \bMM{\text{elec\,(mag)}}{(l-2m+1)} \\[5mm]
&\mspace{10mu} {}+\dsum_{m=0}^{\lfloor (l-2)/2\rfloor}
g_{l,m}\, k^{4m+2}\, \eta^{(l)}_{\text{elec\,(mag)}}\,
\bMM{\text{elec\,(mag)}}{(l-2m)\,\text{\raisebox{0.5ex}{$*$}}}\tc{l-2m-1}   \bMM{\text{elec\,(mag)}}{(l-2m-1)},
\end{array}
\end{equation}
and
\begin{equation}
\begin{array}{lll}
\bigl[\,\bOO{\text{elec}}{(l)}\tc{l-1} \bOO{\text{mag}}{(l)\,\text{\raisebox{0.5ex}{$*$}}}\,\bigr]\! \tc{2} \beps
=\dsum_{m=0}^{\lfloor(l-1)/2\rfloor}
h_{l,m}\, k^{4m}\bigl[\,\eta^{(l)}_{\text{hyb}}\,
\bMM{\text{elec}}{(l-2m)}\tc{l-2m-1}
 \bMM{\text{mag}}{(l-2m)\,\text{\raisebox{0.5ex}{$*$}}}\,\bigr]\! \tc{2} \beps
\end{array}
\end{equation}
with
\begin{eqnarray}
c_{l,m}
&=&
\dfrac{(-1)^m}{4^m}\,\dfrac{l!}{m!}\,\dfrac{\Gamma(l-m+\frac12)}{\Gamma(l+\frac12)\,\Gamma(l-2m)}\,\dfrac{(l-2m)}{l^2}=\dfrac{(-1)^m(l-2m)}l\,d_{l,m},\\
f_{l,m}
&=&
\dfrac{(-1)^m}{4^m}\,\dfrac{l!}{m!}\,\dfrac{\Gamma(l-m+\frac12)}{\Gamma(l+\frac12)\,\Gamma(l-2m)}\,\dfrac{(l-2m+1)(2l-2m+1)}{l(l+1)(2l+1)},
\end{eqnarray}
\begin{eqnarray}
g_{l,m}
&=&
\dfrac{(-1)^m}{4^m}\,\dfrac{l!}{m!}\,\dfrac{\Gamma(l-m+\frac12)}{\Gamma(l+\frac12)\,\Gamma(l-2m)}\,\dfrac{(l-2m)(l-2m-1)}{l(l+1)(2l+1)}, \\
h_{l,m}
&=&
\dfrac{(-1)^m}{4^m}\,\dfrac{l!}{m!}\,\dfrac{\Gamma(l-m+\frac12)}{\Gamma(l+\frac12)\,\Gamma(l-2m)}\,\dfrac{(l-2m)^2}{l^3}.
\end{eqnarray}
\end{subequations}

The decomposition of the optical force into the irrotational and the solenoidal 
terms starts with Eqs.~(\ref{Fmixl}) and (\ref{Fscal}) as well as
the definitions Eq.~(\ref{Omega}). 
To achieve high readability, we assume the particle is located in the regime with $z>0$ and write the incident wave, in which the particle is immersed,
in terms of
the plane wave spectrum representation  \cite{clemmow,mandel,nieto,goodman,pws1,pws2}, 
given below,
\begin{equation} \label{ehinc}
\bs E=\bs E_{\text{inc}}=\ddint \be_{\bs u} \,e^{i k \bs u\.\bs r}\,\td u_x\td u_y \quad \text{and}\quad
\bs H=\bs H_{\text{inc}}=\dfrac1{Z_0}\ddint \bh_{\bs u} \,e^{i k \bs u\.\bs r}\,\td u_x\td u_y,
\end{equation}
where $k$ is the wave number in the background medium,
$$
\bs u=u_x\be_x+u_y\be_y+u_z\be_z,\quad\text{with}\;\;
u_z=\left\{\begin{array}{lll}
\sqrt{1-u_x^2-u_y^2},  & \quad & \text{if } u_x^2+u_y^2\le1 \\
i\sqrt{u_x^2+u_y^2-1}, & \quad & \text{if } u_x^2+u_y^2>1
\end{array}\right. \text{ \;and \;\;} \bs u\.\bs u=1,
$$
whereas the electric and magnetic plane wave spectra $\bs e_{\bs u}$ and $\bs h_{\bs u}$ satisfy
\begin{equation} \label{mynote1}
\bs u\.\bs e_{\bs u}=\bs u\.\bs h_{\bs u}=0,
\quad \bs h_{\bs u}=\bs u\x\bs e_{\bs u}, \quad   \bs e_{\bs u}=-\bs u\x\bs h_{\bs u}.
\end{equation}

With the plane wave spectrum representation Eqs. (\ref{ehinc}), the multiple gradients in (\ref{Omega}) can be written as 
\begin{subequations} \label{mgrad}
\begin{eqnarray}
\bMM{\text{elec}}{(n)}
&=&\dfrac{(ik)^{n-1}}{n}
\dsum_{j=0}^{n-1}\,\ddint
\bs u^{(n-1-j)}\,\be_{\bs u}\,\bs u^{(j)}\,e^{ik\bs u\.\bs r}\td u_x\td u_y, \\
\bMM{\text{mag}}{(n)}
&=&\dfrac{(ik)^{n-1}}{n\,c}
\dsum_{j=0}^{n-1}\,\ddint
\bs u^{(n-1-j)}\,\bh_{\bs u}\,\bs u^{(j)}\,e^{ik\bs u\.\bs r}\td u_x\td u_y,
\end{eqnarray}
\end{subequations}
where $\bs u^{(n)}$ denotes the tensor product of $n$ vectors $\bs u$, viz, $\bs u^{(0)}=1$, $\bs u^{(1)}=\bs u$, $\bs u^{(2)}=\bs u\otimes\bs u$,
and $\bs u^{(3)}=\bs u\otimes\bs u\otimes\bs u$,
with the symbol $\otimes$ representing the tensor product.

After lengthy algebra, the interception parts $\av{\bs F^{\text{\,e}(l)}_{\text{int}}}$ and $\av{\bs F^{\text{\,m}(l)}_{\text{int}}}$
of the optical force involving order $l$ multipoles can be rewritten as
\begin{subequations} \label{FeFmintl}
\begin{equation}
\begin{array}{lcl}
\av{\bs F^{\text{\,e}(l)}_{\text{int}}}
&\!=&\!
\dfrac1{2\,l!}\dsum^{\lfloor\frac{l-1}2\rfloor}_{m=0}
 c_{lm}k^{4m}\, \text{Re}\,
\bigl[\,\gamma^{(l)}_{\text{elec}}\,\bs t^{\,(l-2m)}_{\text{elec}}\,\bigr] \\[4mm]
\av{\bs F^{\text{\,m}(l)}_{\text{int}}}
&\!=&\!
\dfrac1{2\,l!}\dsum^{\lfloor\frac{l-1}2\rfloor}_{m=0}
 c_{lm}k^{4m}\, \text{Re}\,
\bigl[\,\gamma^{(l)}_{\text{mag}}\,\bs t^{\,(l-2m)}_{\text{mag}}\,\bigr]
\end{array}
\end{equation}
where
\begin{eqnarray}
\bs t_{\text{elec}}^{\,\,(n)}
&\equiv&
\bigl[\nabla^{(n)} \bs{E}^*\bigr]\tc{n}\bMM{\text{elec}}{(n)}
\nonumber\\
&=&
\dfrac12\,\Bigl[\nabla D_{\text{ee}}^{(n)}- \nabla\x\bs S_{\text{ee}}^{(n)}-2i\omega\,\text{Re}\,\bs S_{\text{em}}^{(n)}\Bigr] \nonumber \\ & & \mspace{1mu} {}
-\dfrac{(n-1)\omega^2}{2n}\,\Bigl[\nabla D_{\text{mm}}^{(n-1)}-\nabla\x \bs S_{\text{mm}}^{(n-1)}-\dfrac{2i\omega}{c^2}\,\text{Re}\,\bs S_{\text{em}}^{(n-1)}\Bigr], \\[2mm]
\bs t_{\text{mag}}^{\,\,(n)}
&\equiv&
\bigl[\nabla^{(n)} \bs{B}^*\bigr]\tc{n}\bMM{\text{mag}}{(n)}
\nonumber\\
&=&
\dfrac12\,\Bigl[\nabla D_{\text{mm}}^{(n)}- \nabla\x\bs S_{\text{mm}}^{(n)}-\dfrac{2i\omega}{c^2}\,\text{Re}\,\bs S_{\text{em}}^{(n)}\Bigr] \nonumber \\ & & \mspace{1mu} {}
-\dfrac{(n-1)\omega^2}{2n\,c^4}\,\Bigl[\nabla D_{\text{ee}}^{(n-1)}-\nabla\x \bs S_{\text{ee}}^{(n-1)}-2i\omega\,\text{Re}\,\bs S_{\text{em}}^{(n-1)}\Bigr].
\end{eqnarray}
\end{subequations}
The field moments in the reciprocal space are defined as follows.
\begin{equation}\label{fieldmoment}
\begin{array}{lllll}
D_{\text{ee}}^{(n)}
&\!=&\! 
\bigl(\nabla^{(n-1)}\bs E \bigr)\tcj{n}\bigl(\nabla^{(n-1)}\bs E^* \bigr),
      \\[2mm]
D_{\text{mm}}^{(n)}
&\!=&\! 
\bigl(\nabla^{(n-1)}\bs B \bigr)\tcj{n}\bigl(\nabla^{(n-1)}\bs B^* \bigr),
     \\[2mm]
\bs S_{\text{ee}}^{(n)}
&\!=&\!\bigl[ (\nabla^{(n-1)}\bs E)\tcj{n-1}(\nabla^{(n-1)}\bs E^*)\bigr]\tcj{2}\beps,
     \\[2mm]
\bs S_{\text{mm}}^{(n)}
&\!=&\!\bigl[ (\nabla^{(n-1)}\bs B)\tcj{n-1}(\nabla^{(n-1)}\bs B^*)\bigr]\tcj{2}\beps,
     \\[2mm]
\bs S_{\text{em}}^{(n)}
&\!=&\!\bigl[ (\nabla^{(n-1)}\bs E)\tcj{n-1}(\nabla^{(n-1)}\bs B^*)\bigr]\tcj{2}\beps.
\end{array}
\end{equation}
It is noted that $D_{\text{ee}}^{(n)}$ and $D_{\text{mm}}^{(n)}$ are real while
$\bs S_{\text{ee}}^{(n)}$ and $\bs S_{\text{mm}}^{(n)}$ are purely imaginary.
In Eqs. (\ref{fieldmoment}) we have defined the second kind of
multiple tensor contraction, denoted by $\tcj{m}$,
\begin{equation}\label{AtcjB}
\stackrel{\text{\scriptsize$\boldsymbol{\leftrightarrow}$}}
         {\boldsymbol{\mathbb{A}}}^{\text{\raisebox{-1.8ex}{$(n)$}}}
         \!\!\tcj{m}\!\!
\stackrel{\text{\scriptsize$\boldsymbol{\leftrightarrow}$}}
         {\boldsymbol{\mathbb{B}}}^{\text{\raisebox{-1.8ex}{$(n')$}}}=
\mathbb{A}^{(n)}_{\,\textcolor{blue}{k_1\,k_2\,\cdots\,k_m}\,i_{m+1}\,i_{m+2}\,\cdots\,i_n}\,
\mathbb{B}^{(n')}_{\,\textcolor{blue}{k_1\,k_2\,\cdots\,k_m}\,j_{m+1}\,j_{m+2}\,\cdots\,j_{n'}},\qquad
0\le m\le \min[n,n']
\end{equation}
which differs from Eq. (\ref{AtcB}) in that the tensor contraction is made
over the corresponding left most indices, instead of
over the nearest indices, in the two index sequences.
Some simple examples are
\begin{equation}
\begin{array}{llll}
(\bs v\bs w)\tc{2}\beps=\beps\tc{2}(\bs v\bs w)=\bs w\x\bs v, &\quad\text{versus} &\quad
(\bs v\bs w)\tcj{2}\beps=\beps\tcj{2}(\bs v\bs w)=\bs v\x\bs w, \\[2mm]
(\nabla\nabla\bs E)\tc{2}(\nabla\bs B)=(\pdi{i}\pdi{j}E_k)(\pdi{k}B_j)   &\quad\text{versus} &\quad
(\nabla\nabla\bs E)\tcj{2}(\nabla\bs B)=(\pdi{i}\pdi{j}E_k)(\pdi{i}B_j), \\[2mm]
(\nabla\bs B)\tc{2}(\nabla\nabla\bs E)=(\pdi{i}B_j)(\pdi{j}\pdi{i}E_k) &\quad\text{versus} &\quad
(\nabla\bs B)\tcj{2}(\nabla\nabla\bs E)=(\pdi{i}B_{j})(\pdi{i}\pdi{j}E_k). 
\end{array} \label{example}
\end{equation}
Since $\text{Re}\,\bs S_{\text{em}}^{(n)}$ is solenoidal (divergenceless, as can be seen below),
Eq. (\ref{FeFmintl}), together with (\ref{Fmixa}) and (\ref{Fmixb}), concludes the decomposition of the interception (extinction) force.
In the dipolar limit, one has $l=1$, the interception force can be written in terms of
gradients $\nabla D_{\text{ee}}^{(1)}=\nabla |\bs E|^2$
and $\nabla D_{\text{mm}}^{(n)}=\nabla |\bs B|^2$ of field intensity,
the Poynting vector $\bs S_{\text{em}}^{(1)}=\bs E\x\bs B^*$,
and the curl of spin angular momentum density, $\nabla\x(\bs E\x\bs E^*)$ and $\nabla\x(\bs B\x\bs B^*)$,
which has been claimed by many authors as the origin of the optical force (see, e.g., Ref.~\cite{alb}).

The electric and magnetic parts of the recoil force can be worked out to give
\begin{equation}\label{FscaeFscam}
\begin{array}{lll}
\av{\bs F_{\text{rec}}^{\,\text{e}(l)}}
&\!\!=-\dfrac{c_l\,k^{2l+3}}{4\pi\varepsilon_0}\,
\Bigl\{ \dsum_{m=0}^{\lfloor(l-1)/2\rfloor}f_{l,m}k^{4m}\,\text{Im}\bigl[\eta_{\text{elec}}^{(l)}\bs t_{\text{ee}}^{\,(l-2m)\,*}\bigr]
+\dsum_{m=0}^{\lfloor(l-2)/2\rfloor}g_{l,m}k^{4m+2}\,\text{Im}\bigl[\eta_{\text{elec}}^{(l)}\bs t_{\text{ee}}^{\,(l-2m-1)}\bigr]
\Bigr\} \\[5mm]
\av{\bs F_{\text{rec}}^{\,\text{m}(l)}}
&\!\!=-\dfrac{c_l\,k^{2l+3}}{4\pi\varepsilon_0c^2}\,
\Bigl\{ \dsum_{m=0}^{\lfloor(l-1)/2\rfloor}f_{l,m}k^{4m}\,\text{Im}\bigl[\eta_{\text{mag}}^{(l)}\bs t_{\text{mm}}^{\,(l-2m)\,*}\bigr]
+\dsum_{m=0}^{\lfloor(l-2)/2\rfloor}g_{l,m}k^{4m+2}\,\text{Im}\bigl[\eta_{\text{mag}}^{(l)}\bs t_{\text{mm}}^{\,(l-2m-1)}\bigr]
\Bigr\}
\end{array}
\end{equation}
where
\begin{subequations}
\begin{eqnarray}
\bs t^{\,\,(n)}_{\text{ee}}
&\equiv&
\bMM{\text{elec}}{(n)}\tc{n} \bMM{\text{elec}}{(n+1)\,\text{\raisebox{0.5ex}{$*$}}}
=\bMM{\text{elec}}{(n+1)\,\text{\raisebox{0.5ex}{$*$}}}\tc{n} \bMM{\text{elec}}{(n)}  \nonumber \\
&=&
\dfrac12\,\Bigl[\nabla D_{\text{ee}}^{(n)}- \nabla\x\bs S_{\text{ee}}^{(n)}-2i\omega\,\text{Re}\,\bs S_{\text{em}}^{(n)}\Bigr]
\textcolor{blue}{+\dfrac{i\omega}{(n+1)}\,S_{\text{em}}^{(n)}} \nonumber \\
 & & \mspace{1mu} {}
-\dfrac{(n-1)\omega^2}{2\textcolor{black}{(n+1)}}\,\Bigl[\nabla D_{\text{mm}}^{(n-1)}-\nabla\x \bs S_{\text{mm}}^{(n-1)}-\dfrac{2i\omega}{c^2}\,\text{Re}\,\bs S_{\text{em}}^{(n-1)}\Bigr]   \label{tee}\\[2mm]
\bs t^{\,\,(n)}_{\text{mm}}
&\equiv&
\bMM{\text{mag}}{(n)}\tc{n} \bMM{\text{mag}}{(n+1)\,\text{\raisebox{0.5ex}{$*$}}}
=\bMM{\text{mag}}{(n+1)\,\text{\raisebox{0.5ex}{$*$}}}\tc{n} \bMM{\text{mag}}{(n)}  \nonumber\\
&=&
\dfrac12\,\Bigl[\nabla D_{\text{mm}}^{(n)}- \nabla\x\bs S_{\text{mm}}^{(n)}-\dfrac{2i\omega}{c^2}\,\text{Re}\,\bs S_{\text{em}}^{(n)}\Bigr]
\textcolor{blue}{+\dfrac{i\omega}{(n+1)c^2}\,S_{\text{em}}^{(n)\,*}} \nonumber \\
 & & \mspace{1mu} {}
-\dfrac{(n-1)\omega^2}{2\textcolor{black}{(n+1)c^4}}\,\Bigl[\nabla D_{\text{ee}}^{(n-1)}-\nabla\x \bs S_{\text{ee}}^{(n-1)}-2i\omega\,\text{Re}\,\bs S_{\text{em}}^{(n-1)}\Bigr] \label{tmm}
\end{eqnarray}
\end{subequations}
with $c_l={(l+2)}/[{(l+1)!(2l+3)!!}]$.

Finally, the hybrid term $\av{\bs F^{\text{\,x}(l)}_{\text{rec}}}$ of the recoil force, given by (\ref{Fscalx}), can be cast, for our purpose, into
\begin{equation} \label{Fscaxl}
\av{\bs F_{\text{rec}}^{\,\text{x}(l)}}
=\dfrac{Z_0}{4\pi}\,\dfrac{k^{2l+2}}{l\,l!(2l+1)!!}
 \dsum_{m=0}^{\lfloor(l-1)/2\rfloor}h_{l,m}k^{4m}\,\text{Re}\bigl[\eta_{\text{hyb}}^{(l)}\bs t_{\text{em}}^{\,(l-2m)}\bigr].
\end{equation}
where, written in a symmetric form with respect to the electric and magnetic fields (possessing so called ``electric-magnetic
democracy'' \cite{berry2009}),
\begin{eqnarray}
\bs t^{\,\,(n)}_{\text{em}}
&\!\! \equiv &\!\!
\bigl[\,\bMM{\text{elec}}{(n)}\tc{n-1} \bMM{\text{mag}}{(n)\,\text{\raisebox{0.5ex}{$*$}}}\,\bigr]\tc{2}\beps \nonumber \\
&\!\!=&\!\!
       -\dfrac{i(n-1)\omega}{2n\,c^2}\,\nabla\x\bigl[\bs S_{\text{ee}}^{(n-1)}+c^2 \bs S_{\text{mm}}^{(n-1)}\bigr]
+\dfrac{i(n-1)(n-2)\omega k^2}{2n^2c^2}\,\nabla\x\bigl[\bs S_{\text{ee}}^{(n-2)}+c^2 \bs S_{\text{mm}}^{(n-2)}\bigr]
-\textcolor{blue}{\dfrac1n\,\bs S_{\text{em}}^{(n)}}
\nonumber \\[1mm] &\!\!&\!\! \mspace{-15mu} {}
          -\dfrac{(n-1)}{2n}\,\nabla\x\bigl[\bs G_{\text{em}}^{(n-1)   }- \bs G_{\text{me}}^{(n-1)\,*}\bigr]
+\dfrac{(n-1)(n-2)k^2}{2n^2}\,\nabla\x\bigl[\bs G_{\text{em}}^{(n-2)\,*}- \bs G_{\text{me}}^{(n-2)   }\bigr]
+\textcolor{blue}{\dfrac{(n-1)k^2}{n^2}\,\bs S_{\text{em}}^{\,(n-1)\,*}} \label{tem}\mspace{30mu}
\end{eqnarray}
The two new field moments, $\bs G_{\text{em}}^{(n)}$ and $\bs G_{\text{me}}^{(n)}$, in the reciprocal space are given by
\begin{equation}
\bs G_{\text{em}}^{(n)} =[\nabla^{(n-1)}\bs E]\tcj{n}[\nabla^{(n)}\bs B^*], \qquad
\bs G_{\text{me}}^{(n)}=[\nabla^{(n-1)}\bs B]\tcj{n}[\nabla^{(n)}\bs E^*].
\end{equation}
In the dipolar limit, with $l = 1$, the hybrid term depends only on the complex Poynting vector
$\bs S_{\text{em}}^{(1)}=\bs E\x\bs B^*$, which is interpreted as a lateral optical force
resulting from the Belinfante spin momentum \cite{beksh2015,belinf1940} in two-wave interference.

The recoil force, given by (\ref{FscaeFscam} - \ref{tem}), depends on $\bs S_{\text{em}}^{(n)}$, which is neither irrotational
nor solenoidal.
So we need to decompose $\bs S_{\text{em}}^{(n)}$ into a gradient and curl parts. To this end, we write
\begin{equation}\label{decSem}
\bs S_{\text{em}}^{(n)}=-\nabla\varphi_s^{(n)}+\nabla\x\bs\psi_s^{(n)}.
\end{equation}
It is worked out that
\begin{subequations}
\begin{eqnarray}
\varphi_s^{(n)}
&=&-\dfrac{ik^{2n}}{2\omega}\,\dsum_{m=n}^{\infty} \dfrac{1}{k^{2m}}
   \bigl[D_{\text{ee}}^{(m)}-c^2D_{\text{mm}}^{(m)}\bigr], \\
\bs\psi_s^{(n)}
&=&-\dfrac{k^{2(n-1)}}{2}\,\dsum_{m=n}^{\infty} \dfrac{1}{k^{2m}}
   \Bigl[ \bs G_{\text{em}}^{(m)} - \bs G_{\text{me}}^{(m)*} \Bigr],
   \end{eqnarray}
\end{subequations}
which result in
\begin{equation} \label{Sem}
\bs S_{\text{em}}^{(n)}= 
-\dfrac{k^{2n-2}}2\,\nabla\x\dsum_{m=n}^{\infty}\dfrac1{k^{2m}}\bigl[\bs G^{(m)}_{\text{em}}-\bs G^{(m)*}_{\text{me}} \bigr]
+\dfrac{ik^{2n}}{2\omega}\,\nabla\dsum_{m=n}^{\infty}\dfrac1{k^{2m}}\bigl[D^{(m)}_{\text{ee}}-c^2 D^{(m)}_{\text{mm}} \bigr].
\end{equation}
Since $D^{(n)}_{\text{ee}}$ and $D^{(n)}_{\text{mm}}$ are real,
the real part Re$\,\bs S_{\text{em}}^{(n)}$ is divergenceless, as stated in the text after Eq. (\ref{example}).
Equations (\ref{FscaeFscam} - \ref{tem}), together with (\ref{Sem}), (\ref{Fscaa}), and (\ref{Fscab}), constitute the decomposition of the recoil force
into a conservative and a nonconservative parts.


\end{document}